\documentclass{iau} 
\usepackage{graphicx}

\title[Missing flux of SiO maser at 7\,mm in IRC+10011] %% give here short title %%
{Missing flux in VLBI observations of SiO maser at 7\,mm in IRC+10011}

\author[Desmurs$^1$, J.-F. et al.] %% give here short author list %%
{Desmurs$^1$, J.-F., Alcolea$^1$, J., Bujarrabal$^1$, V., Colomer$^{1,2}$, F. 
\and Soria-Ruiz$^1$, R.}

   \affiliation{$^1$Observatorio Astronómico Nacional (OAN/IGN), Spain\\[\affilskip]
     $^2$JIVE, The Netherlands}

\pubyear{2017}
\volume{336}  %% insert here IAU Symposium No.
\jname{Astrophysical Masers: Unlocking the Mysteries of the Universe}
\editors{A. Tarchi, M.J. Reid \& P. Castangia, eds.}
\begin{document}

\maketitle

\begin{abstract}
VLBI observations of SiO masers recover at most 40-50\% of the total
flux obtained by single dish observations at any spectral channel. Some
previous studies seems to indicate that, at least, part of the lost
flux is divided up into many weak components rather than in a large
resolved emission area. Taking benefit of the high sensitivity and
resolution of the HSA, we investigate the problem of the missing flux
in VLBI observations of SiO maser emission at 7\,mm in the AGB stars and
obtain a high dynamic range map of IRC+10011. We conclude that the
missing flux is mostly contained in many very weak maser components.
\end{abstract}

\firstsection % if your document starts with a section,
\section{Introduction}

VLBI observations of SiO masers are providing extremely valuable
information on the inner circumstellar shells around AGB stars, the
regions where dust grains are not yet formed and mass ejection
originates, after a complex pulsational dynamics. These data are also
very useful to understand the pumping mechanisms responsible for this
widespread emission in AGB envelopes. The $J$=1--0 maser lines (in the $v$=1
and $v$=2 vibrationally excited states), at 7\,mm wavelength,
systematically yield ring-like flux distributions, with diameters of
about 10$^{14}$\,cm (equivalent to a few stellar radii, see
\cite[Diamond et al. 1994]{Diamond_etal94},
\cite[Desmurs et al. 2000]{Desmurs_etal00}).

One of the main problems that persists in the study of the
circumstellar SiO masers is the significant amount of flux lost when
long baseline interferometry observations are performed. For 7\,mm
lines, up to about one half of the line emission is usually lost, as it
is also the case at 3\,mm (see \cite[Colomer et
  al. 2017]{Colomer_etal17}), a problem that is not present in VLA
observations. This missing flux could be due to over-resolution,
i.e. when the emission is produced on larger scales than those
corresponding to the shortest projected baselines of the
array. However, another possible explanation could be that this missing
flux, or at least part of it, consists of a multitude of compact but
weak undetected maser components (at the noise level of the resulting
map).

\begin{figure}[t]
% \vspace*{-2.0 cm}
\begin{center}
  \includegraphics[width=0.24\textwidth]{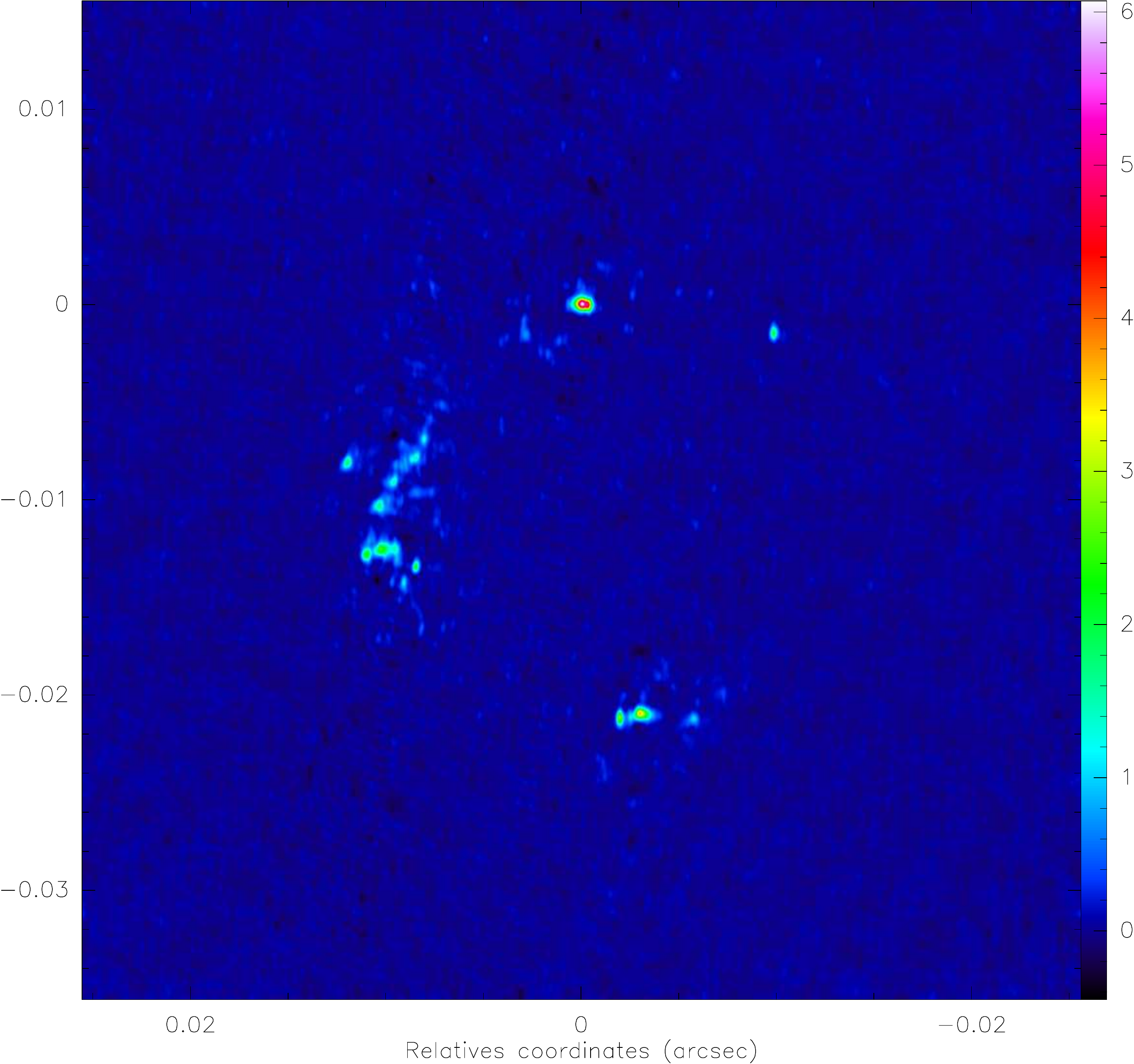}
  \includegraphics[width=0.24\textwidth]{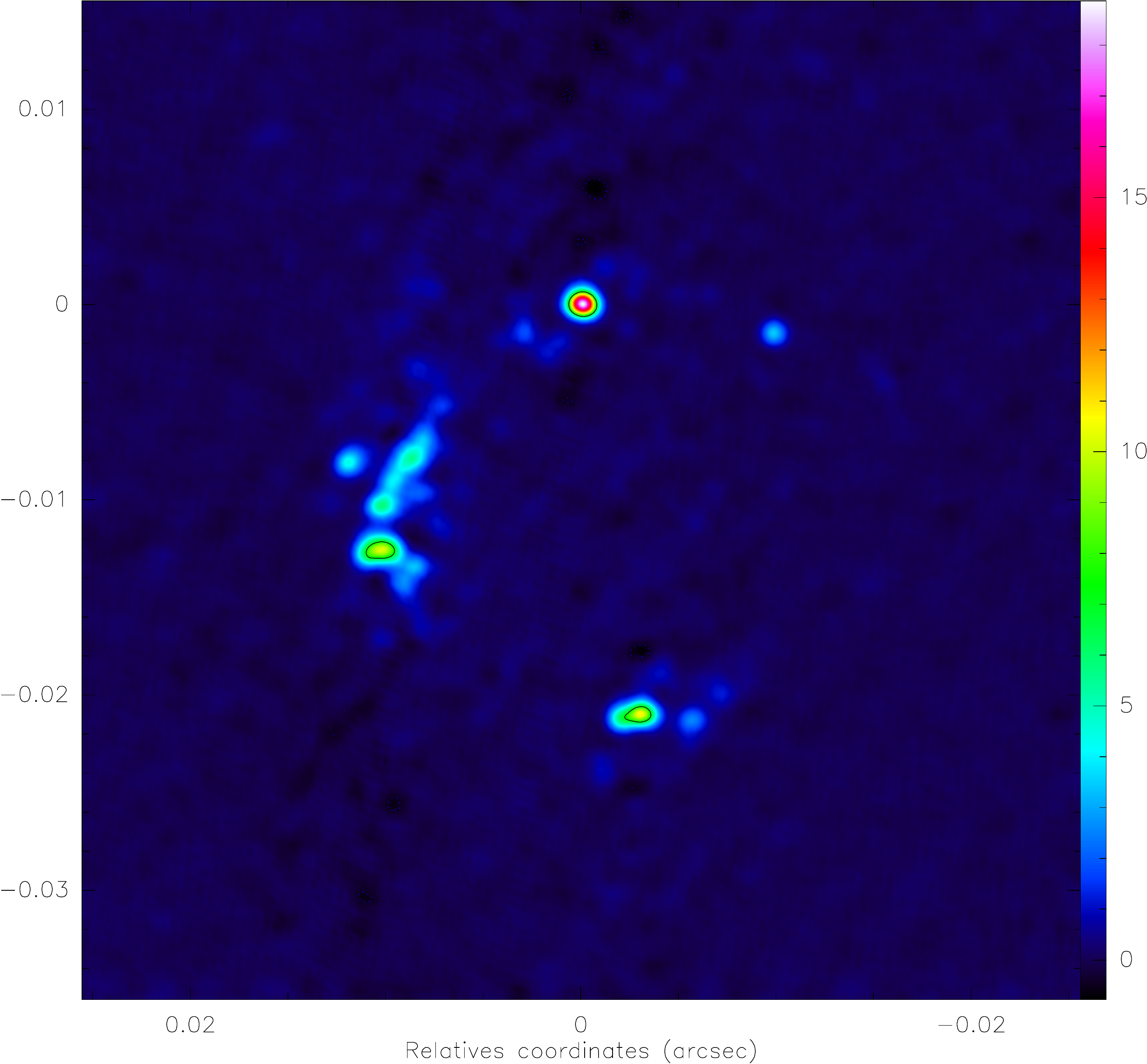}
  \includegraphics[width=0.24\textwidth]{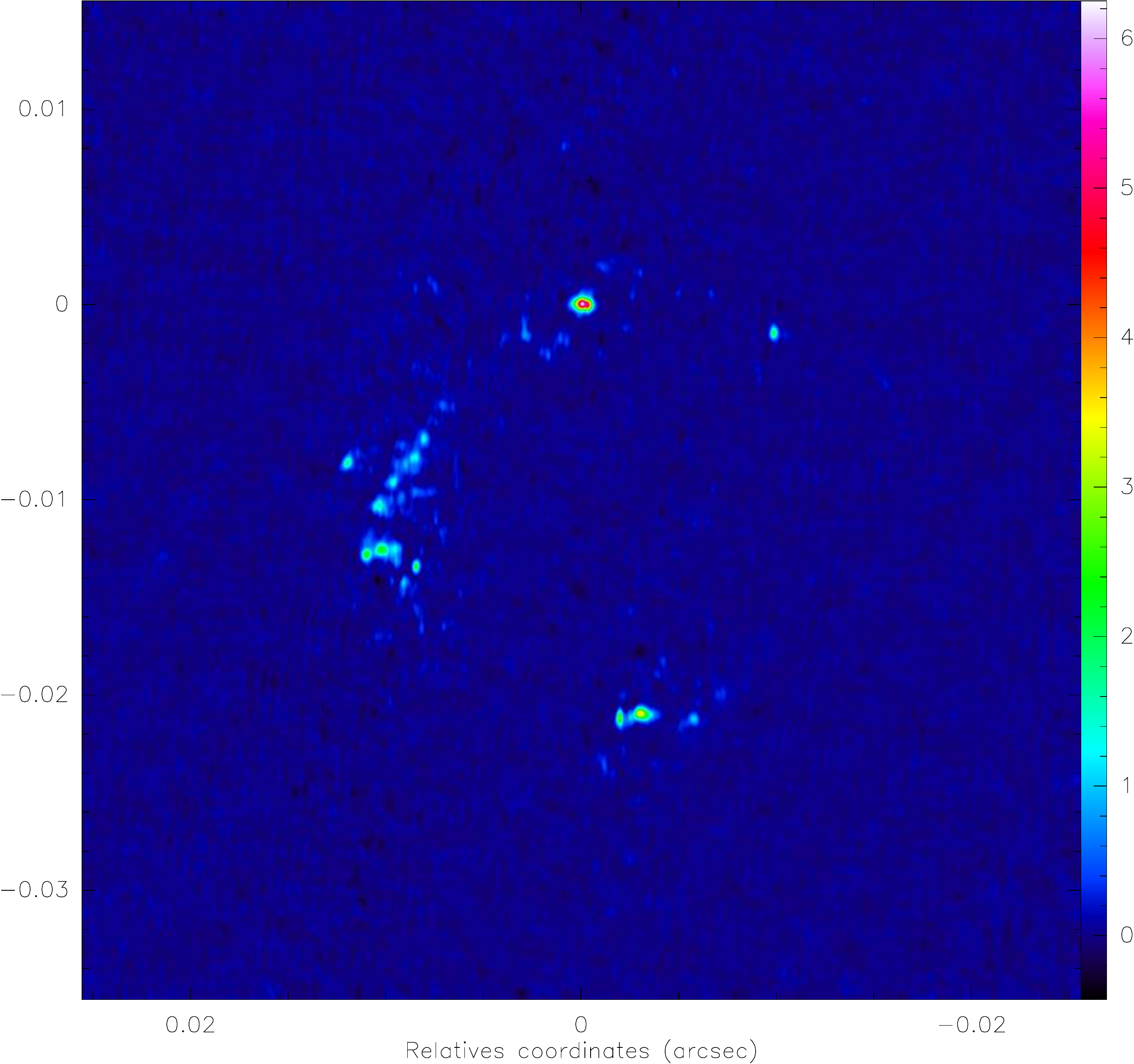}
  \includegraphics[width=0.24\textwidth]{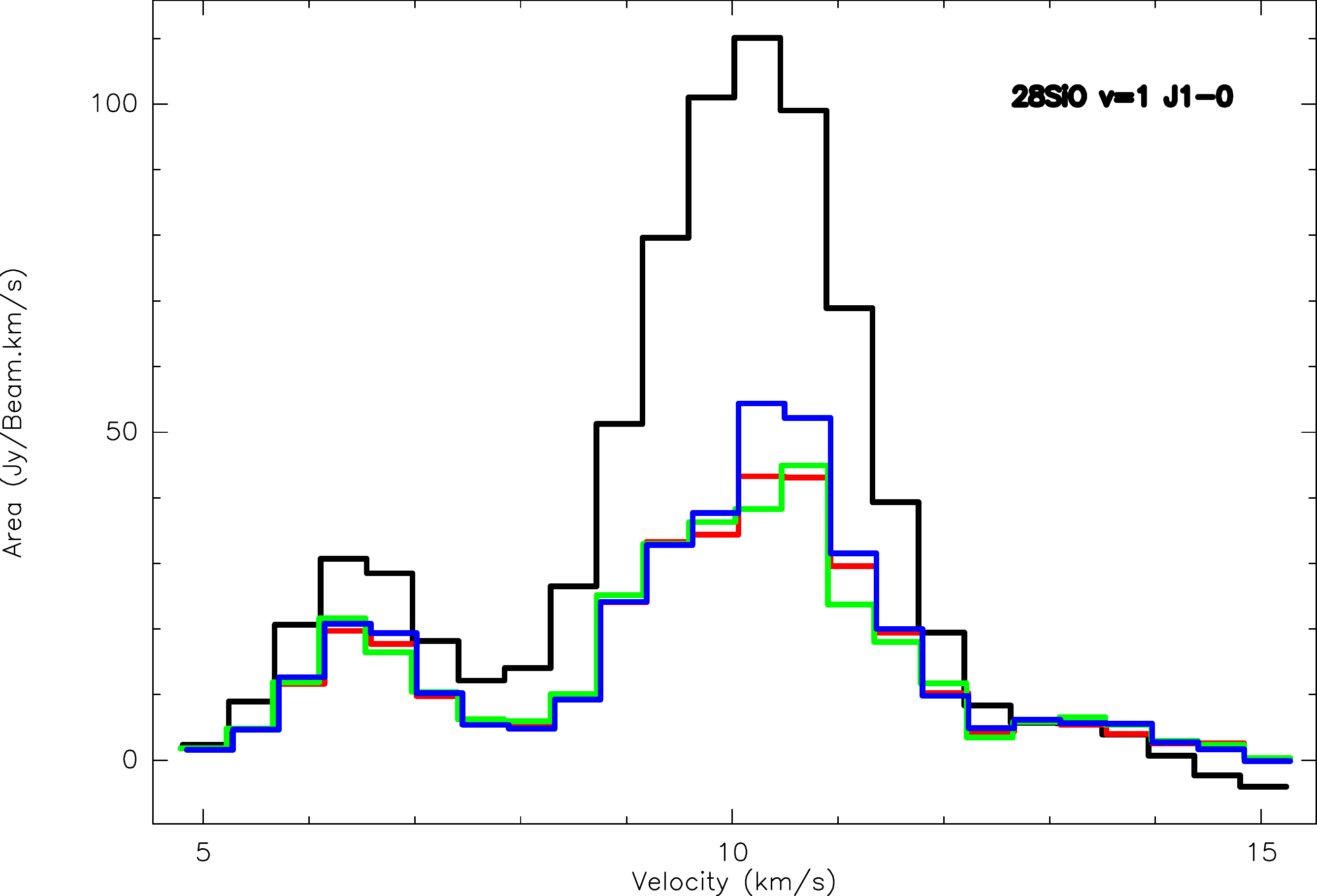}

  \includegraphics[width=0.24\textwidth]{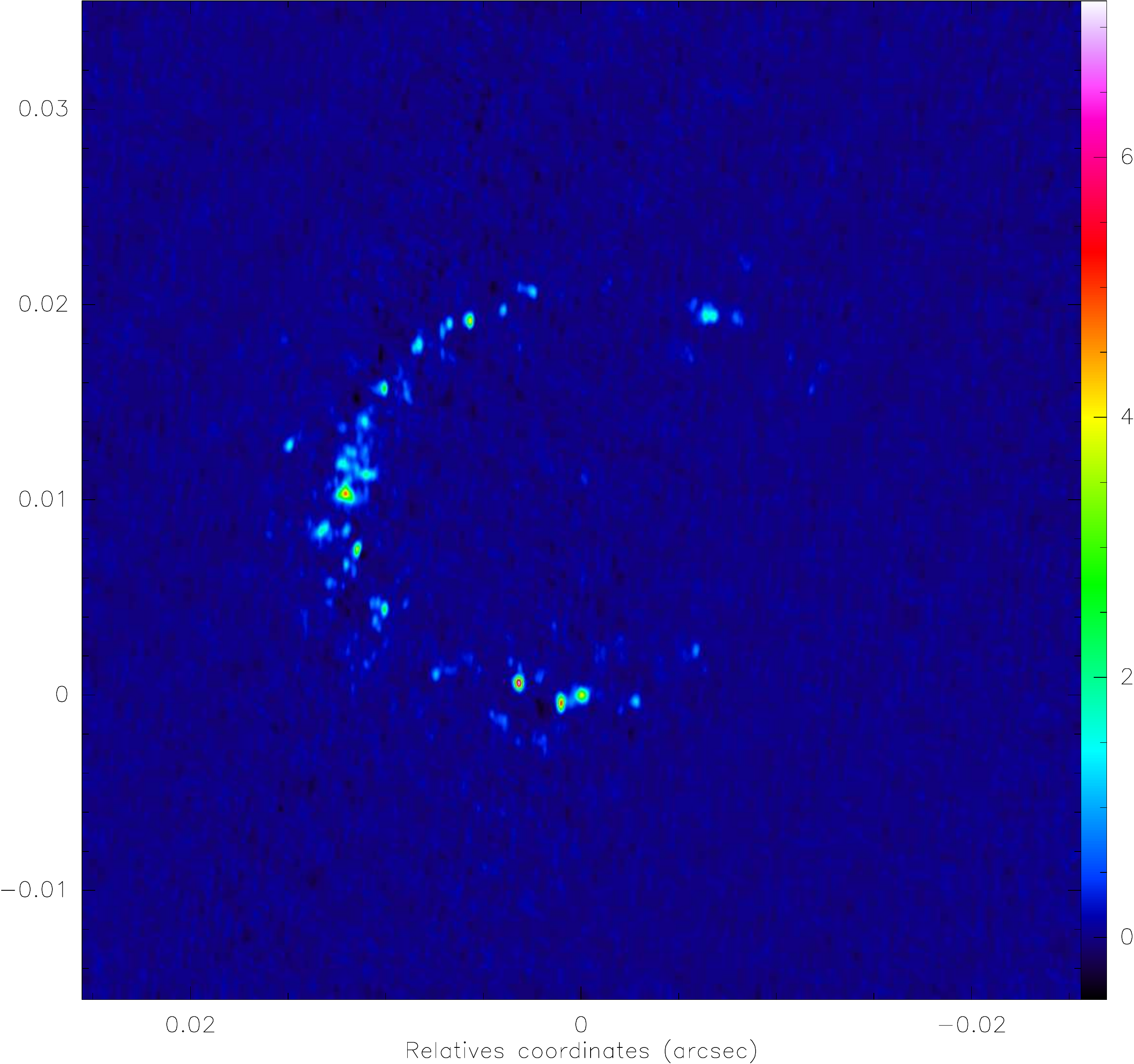}
  \includegraphics[width=0.24\textwidth]{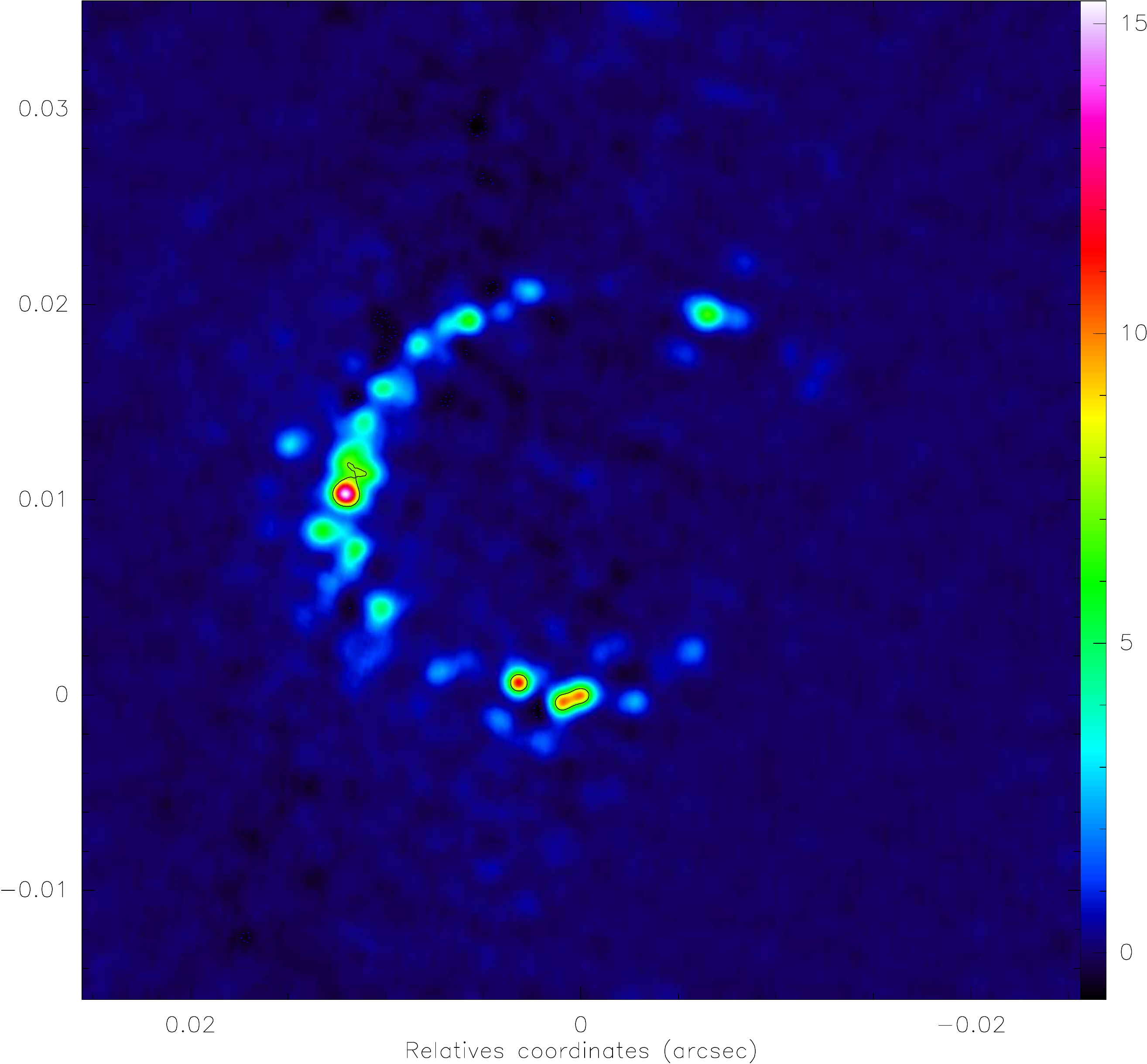}
  \includegraphics[width=0.24\textwidth]{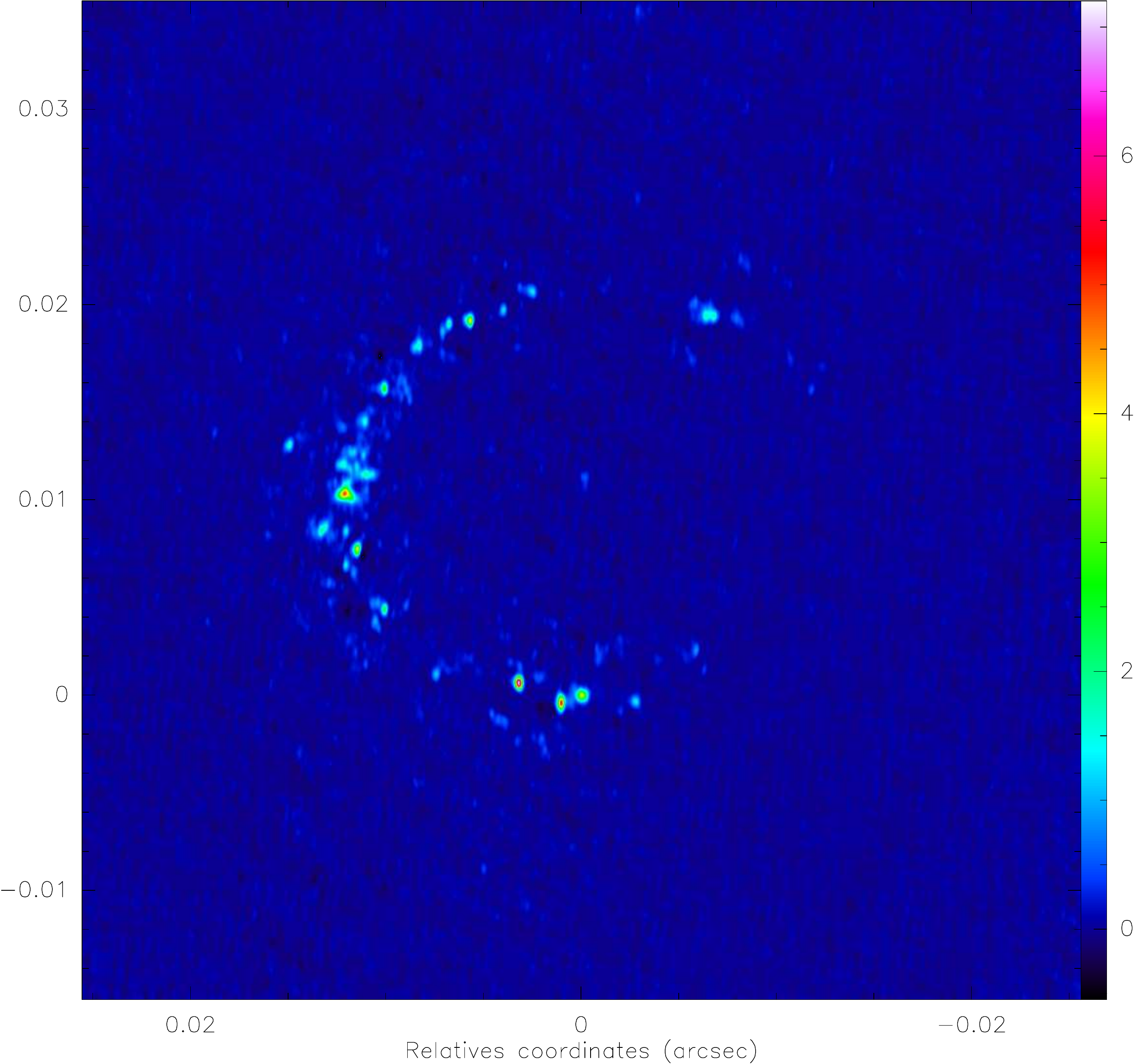}
  \includegraphics[width=0.24\textwidth]{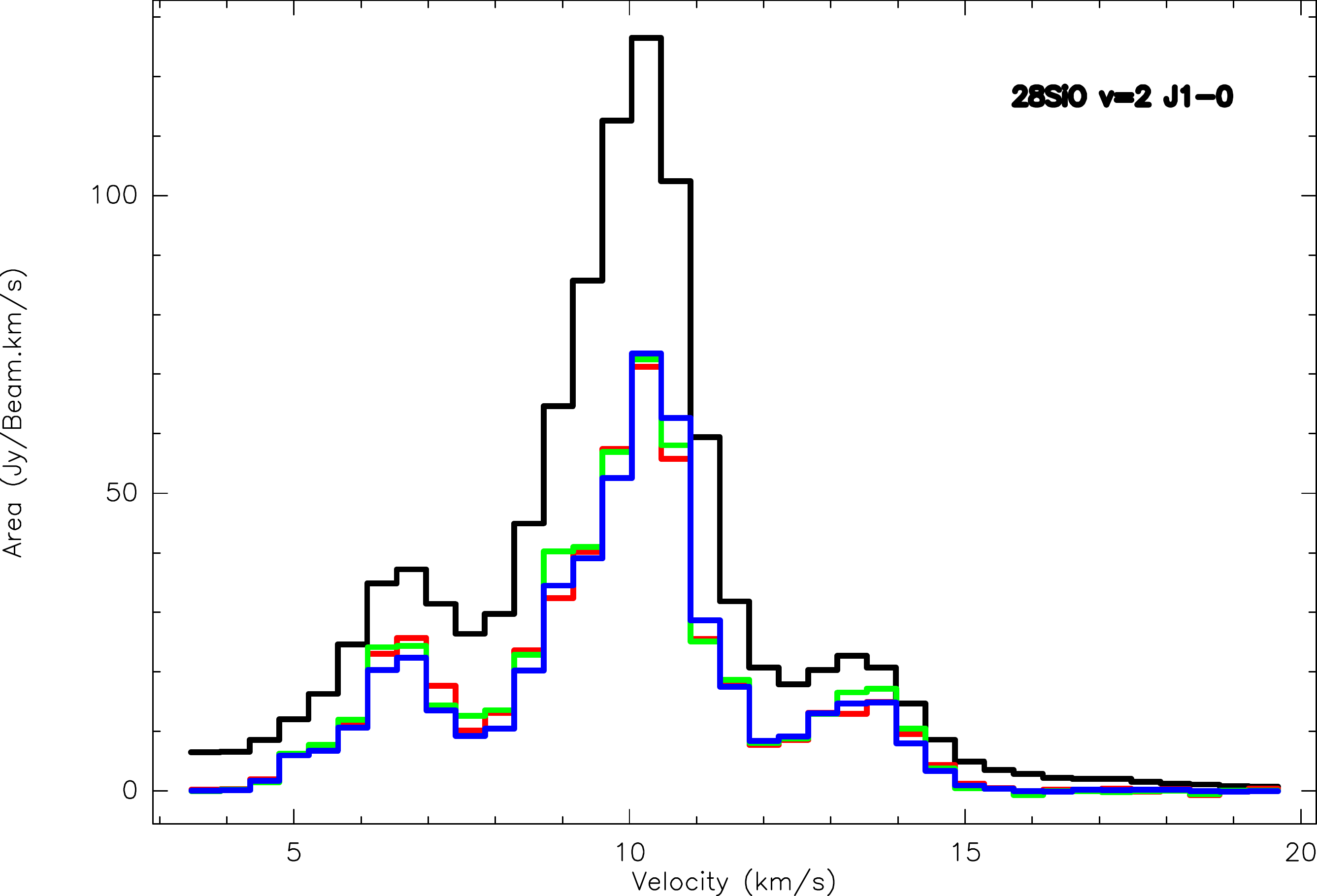}
% \vspace*{-1.0 cm}
 \caption{{\it Top} $^{28}$SiO $v$=1, $J$=1--0 (at 43.122\,GHz)
   transition and {\it Bottom}
   $^{28}$SiO $v$=2, $J$=1--0 (at 42.820\,GHz) transition. From left to right,
   maps of the two maser transitions obtained respectively  with, case
   A, the full HSA array with full sensitivity and spatial resolution
   (baselines up to $\sim$10500\,km, restoring beam 0.2\,mas), case B, with a
   subset of antennas forming a very sensitive short array (with
   baselines $<$2500\,km) and, case C, with all HSA antennas (and full
   sensitivity) but with a degraded restoring beam of 1\,mas (low
   resolution). At right, flux density comparison between the
   autocorrelation flux intensity of the reference antenna used for the
   flux calibration (black line) and the integrated flux recovered in
   the maps in case A (red line), in case B (green line), and case C
   (blue line).}
   \label{fig1}
\end{center}
\end{figure}

\section{Observations}

To check if part of the missing flux is contained in many very weak
maser components (see \cite[Soria-Ruiz et al. 2004]{Soria_etal04}) or
not, we took advantage of HSA capabilities at 7\,mm that give a better
UV-coverage, higher sensitivity and higher resolution. We observed in
dual circular polarization with a velocity resolution (i.e. channel
width) of 0.2\,km/s and a total velocity coverage of about
55\,km/s. Using all VLBA antennas, the VLA, the GBT and Effelsberg, we
obtained maps of IRC+10011 of the two $^{28}$SiO transitions $v$=1 and $v$=2,
$J$=1--0 with a high spatial resolution and a high dynamic range (see
Fig.\,\ref{fig1}).

\section{Preliminary results}

Our preliminary results tend to show very similar results for the
proportion of missing flux measured in these observations and in
previous works. About half of the flux is still missing! The high
sensitivity, we reach an rms of about 5\,mJy/beam per channel for
$^{28}$SiO $v$=2, $J$=1--0, and high resolution of HSA ($\sim$0.2\,mas)
do not allow us to significantly recover a higher percentage of
flux. Moreover, either using the full spatial resolution of HSA with
baseline of up to 10500\,km or a compact array with baselines shorter
than 2500\,km (including short baseline highly sensitive VLBA-PT/VLA),
do not significantly change this result. Even degrading the resolution
(using a restoring beam 5 times larger), the small flux increase
measured in a couple of channels (@ 10-11\,km/s) for the $v$=1 map and
corresponding to the arc like structure seen on the east side of the
map is not significant, there is no difference in the recovering
flux. Our main idea to explain these results is that the missing flux
must be spread in a multitude of weak components undetected in our
observations.


\begin{thebibliography}{}

\bibitem[Colomer \etal\ (2017)]{Colomer_etal17}
 {Colomer, F., Desmurs, J.-F., Bujarrabal, V.  et al.} 2017, \textit{HSA
  IX, proc. SEA  held on jul 18-22, 2016 in Bilbao}, p361-366

\bibitem[Desmurs  \etal\ (2000)]{Desmurs_etal00}
 {Desmurs, J.-F., Bujarrabal V., Lindqvist M., et al.} 2000, \textit{A\&A}, 565, 127

\bibitem[Diamond \etal\ 1994]{Diamond_etal94}
 {Diamond, P. J., Kemball, A. J., Junor, W., et al.} 1994, \textit{ApJ}, 430, L61
  
\bibitem[Soria-Ruiz \etal\ (2004)]{Soria_etal04}
 {Soria-Ruiz, R., Alcolea, J., Colomer, F.  et al.} 2004, \textit{A\&A}, 426, 131

\end{thebibliography}
\end{document}